\documentstyle[12pt]{article}
\newcommand{\be}{\begin{equation}}
\newcommand{\ee}{\end{equation}}
\newcommand{\bea}{\begin{eqnarray}}
\newcommand{\eea}{\end{eqnarray}}
\def\pd{\partial}
\makeatletter
\def\eqnarray{\stepcounter{equation}\let\@currentlabel=\theequation
\global\@eqnswtrue
\global\@eqcnt\z@\tabskip\@centering\let\\=\@eqncr
$$\halign to \displaywidth\bgroup\@eqnsel\hskip\@centering
  $\displaystyle\tabskip\z@{##}$&\global\@eqcnt\@ne
  \hfil$\displaystyle{\hbox{}##\hbox{}}$\hfil
  &\global\@eqcnt\tw@ $\displaystyle\tabskip\z@
  {##}$\hfil\tabskip\@centering&\llap{##}\tabskip\z@\cr}
\@addtoreset{equation}{section}
  \def\theequation{\thesection.\arabic{equation}}
\makeatother
\begin{document}
\begin{titlepage}
\date{\rightline {DTP/93/39}
\rightline{September 1993}}\title{Integrable Generalisations of the
2-dimensional Born Infeld Equation.}

\author{D.B. Fairlie and J.A. Mulvey\\
 { \it University of Durham,}\\
 South Road, Durham, DH1 3LE, UK}

\maketitle

\begin{abstract}
The Born-Infeld equation in two dimensions is generalised to higher dimensions
whilst retaining Lorentz Invariance and complete integrability. This
generalisation retains homogeneity in second derivatives of the field.

\end{abstract}
\end{titlepage}
\section{Introduction}
There are many nonlinear integrable equations now known in more than 2
dimensions; the Davy-Stewartson equation \cite{dav}, the
Kadomtsev-Petviashvilli equation \cite{KP} and the Konopelchenko-Rogers
equation \cite{rogers} being well known examples. Most such equations including
those particular ones suffer from
a lack of Lorentz invariance, so are inappropriate as field theoretic models
with particle-like solutions. In the case of 1+1 dimensions, there are two
well known integrable nonlinear equations which are Lorentz invariant which
admit localised solutions and have been used to model fundamental particles;
the Sine-Gordon equation \cite{skyrme} and the Born-Infeld equation \cite
{born}, which
in light-cone co-ordinates in 1+1 dimensions is given by
\be
(\frac{\pd\phi}{\pd x})^2\frac{\pd^2\phi}{\pd t^2}+(\frac{\pd\phi}{\pd
t})^2\frac{\pd^2\phi}{\pd x^2}- (\lambda+2\frac{\pd\phi}{\pd
x}\frac{\pd\phi}{\pd t})\frac{\pd^2\phi}{\pd x\pd t}=0.
\label{2}
\ee
The parameter $\lambda$, if non zero, can be scaled to unity. This equation
is known to be integrable \cite{whit}\cite{bar}\cite {nutku}. In this article
we should like to view it as an integrable deformation of the Bateman equation,
which corresponds to (\ref{2}) when $\lambda=0$, in analogy with the integrable
deformations of Conformal Field Theories proposed by Zamolodchikov \cite{zam}.
The properties of covariance (if $\phi(x,t)$ is a solution to the Bateman
equation so is any function of $\phi(x,t)$) and of the existence of an infinite
class of inequivalent  Lagrangian densities from which the equation may be
derived \cite{fai} are lost, but that of integrability is retained. Also the
 curious property of the Bateman equation that it is form invariant under
arbitrary linear transformations of the co-ordinates $(x,t)$, which means that
the Bateman equation is `signature blind' is replaced by Poincar\'e invariance.

In some recent work \cite{fai},\cite{dbf},\cite{fab}, one of us  and his
collaborators have proposed a class of completely
integrable models which we call Universal Field Equations, of which the Bateman
equation is the 2 dimensional prototype. This equation, describing a theory in
$d$ dimensions may be written in the following equivalent forms:
\be
\det\pmatrix{0& \frac{\pd \phi}{\pd x_j}\cr
         \frac{\pd \phi}{\pd x_i}&\frac{\pd^2 \phi}{\pd x_i\pd x_j}\cr}=0.
\label{univ}
\ee
or ${\rm Trace}(GA)=0$, where the matrix $G$  has components
\be
G_{ij}= \frac{\pd \phi}{\pd x_i} \frac{\pd \phi}{\pd x_j}
\ee
and
\be
A_{ij}={\rm Adj}\Bigl( \frac{\pd^2 \phi}{\pd x_k\pd x_l}\Bigr)_{ij}=\det\Bigl(
\frac{\pd^2 \phi}{\pd x_k\pd x_l}\Bigr)\Bigl( \frac{\pd^2 \phi}{\pd x_k\pd
x_l}\Bigr)_{ij}^{-1}
\ee
is the adjugate matrix of $M_{ij}=\frac{\pd^2 \phi}{\pd x_i\pd x_j}$. Under a
Lorentz transformation generated by the matrix $\Lambda$, both $G$ and $A$
transform
in the same way as
\be
G'=\Lambda^{-1}G\Lambda,\quad A'=\Lambda^{-1}A\Lambda.
\label{trans}
\ee
 From the second form of equation (\ref{univ}) it is easy to see that this
equation is invariant under Lorentz transformations. How may this equation be
deformed in such a way as to retain the properties of Lorentz invariance and
integrability? The only other matrix available to us which transforms according
to (\ref{trans}) is the metric tensor $\eta_{ij}$, and thus the only candidate
with this structure since $G$ is idempotent is the equation
\be
\sum_{i,j}A_{ij}(G_{ji}+f({\rm Trace}(\eta G))\eta_{ji})
\label{univ2}
\ee
where $f$ is an arbitrary function of the quadratic Lorentz invariant
constructed from $\frac{\pd \phi}{\pd x_i}$. We are excluding the possibility
of an explicit dependence upon $\phi$, and retain the homogeneity in second
derivatives.  In the following considerations we shall take $f$ to be  a
constant, $\lambda$, generalising (\ref{2}), with a brief mention of the more
general case when we discuss linearization. Using the results of \cite{fab}, we
shall show how both of these forms give rise to equations linearizable by a
Legendre Transform. It is crucial for the proof of integrability given below
that the second order derivatives enter through only components of the adjugate
matrix which appear linearly.  The only other possibility, tractable by this
method is for an additional linear dependence upon det$M$.
 Other Lorentz invariant equations could be constructed using
powers of the matrix $M$, including the Barbashov-Chernikov generalisation of
two dimensional Born-Infeld equation to higher dimensions \cite{bar},  but
there is no incontrovertible evidence for the integrability of such equations,
despite some hopeful remarks in the literature \cite{hoppe}. Note tahat neither
this
generalisation nor the generalisation presented here are equivalent to the
original Born-Infeld equation,\cite{born}, but the terminology seems to be
establihed for the two dimensional situation.
The 3 dimensional version of these equations is a candidate for an alternative
theory of strings, as it describes the motion of a surface $\phi(x_1,x_2,x_3)$,
the world sheet of a string. The case when $\lambda=0$, which just describes
developable surfaces \cite{pleb} was treated in \cite{dbf}. The characteristic
property of such surfaces is that they contain straight lines.

We are thus led to essentially three forms of nonlinear field equations in four
dimensions  which are Lorentz invariant and are integrable; the well known
system of Self Dual Yang Mills equations and its supersymmetric extensions
\cite{yang},\cite{ward} , the relativistic string equations \cite{shaw}
\cite{dbf2} which are a bit of a cheat as the base space is two dimensional and
the equations
proposed here, which are directly related to linear equations through the
Legendre Transform.

The paper is organised as follows.
In the next  section a new derivation of some well known properties of the
Born-Infeld equation in 2 dimensions is given. The proof of integrability goes
back at least to \cite{bar} and is discussed further in
\cite{nutku1},\cite{nutku}, but we shall present a
slightly different version for completeness. Section 3  briefly reviews the
Legendre Transform method, and shows how the
Lorentz invariant deformations of  the Universal Field Equation given by
(\ref{2}) may be solved implicitly. The Lagrangian for this equation
(\ref{univ2}) is constructed. Unlike the case of the Universal Field Equation,
it is unique, thus resolving an ambiguity as to how such theories night be
quantised by the Feynman path integral method. On the other hand,
the equation of motion also follows from the same iterative procedure described
in \cite{fai}.

\section{The Born Infeld Equation}

The Born Infeld equation in light cone co-ordinates in 1+1 dimension is given
by
\be
(\frac{\pd\phi}{\pd x})^2\frac{\pd^2\phi}{\pd t^2}+(\frac{\pd\phi}{\pd
t})^2\frac{\pd^2\phi}{\pd x^2}- (\lambda+2\frac{\pd\phi}{\pd
x}\frac{\pd\phi}{\pd t})\frac{\pd^2\phi}{\pd x\pd t}=0.
\label{1}
\ee
The parameter $\lambda$, if non zero can be scaled to unity. This equation can
be viewed as an integrable deformation of the Bateman equation, which
corrresponds to $\lambda = 0$, in analogy with the integrable deformations of
Conformal Field Theories proposed by Zamolodchikov \cite{zam}.

This equation can be written as a first order equation in a similar manner to
the Bateman equation with the help of the two independent roots $u_1,\  u_2$ of
the quadratic equation for the characteristics \cite{whit};
\be
(\frac{\pd\phi}{\pd x})^2u^2- (\lambda+2\frac{\pd\phi}{\pd x}\frac{\pd\phi}{\pd
t})u+(\frac{\pd\phi}{\pd t})^2=0.
\ee
The roots of this equation are
\be
u_{1\atop2}=\frac{\lambda+2\frac{\pd\phi}{\pd x}\frac{\pd\phi}{\pd t}\pm
\sqrt{\lambda^2+4\lambda\frac{\pd\phi}{\pd x}\frac{\pd\phi}{\pd
t}}}{2(\frac{\pd\phi}{\pd x})^2}
\ee
The Born Infeld equation can then be written in either of two forms;
\bea
\frac{\pd u_1}{\pd t}=&u_2\frac{\pd u_1}{\pd x}\\
\frac{\pd u_2}{\pd t}=&u_1\frac{\pd u_2}{\pd x}
\eea
 These equations possess an infinite number of conservation laws; it is easy to
verify that
\bea
\frac{\pd}{\pd t}(u_1+u_2)=&\frac{\pd }{\pd x}(u_1u_2)\\
\frac{\pd}{\pd t}(u_1^2+u_1u_2+u_2^2)=&\frac{\pd }{\pd x}(u_1u_2(u_1+u_2))\\
\frac{\pd}{\pd t}(u_1^3+u_1^2u_2+u_1u_2^2+u_2^3)=&\frac{\pd }{\pd
x}(u_1u_2(u_1^2+u_1u_2+u_2^2))\\
\dots&{\rm etc.} \nonumber
\eea
In fact, if $S_n$ denotes the symmetric polynomial of $n$th degree in  $u_1,\
u_2$, then the general conservation law is
\be
\frac{\pd}{\pd t}S_n=\frac{\pd }{\pd x}(u_1u_2S_{n-1})
\ee
This is easily proved using $S_n=u_i^n+u_1S_{n-1}$ and induction.
The general solution of the equations for $u_1,\ u_2$  is an implicit one;
The roles of dependent and independent variables may be interchanged to
give
\bea
\frac{\pd x}{\pd u_2}=&-u_2\frac{\pd t}{\pd u_2} \nonumber \\
\frac{\pd x}{\pd u_1}=&-u_1\frac{\pd t}{\pd u_1}
\label{hier}
\eea
with the solution,
\bea
x&=&f(u_1)-u_1f'(u_1)+g(u_2)-u_2g'(u_2) \nonumber \\
t&=&f'(u_1)+g'(u_2)
\label{hodo}
\eea
where $f,\ g$ are arbitrary functions and a prime denotes differentiation with
respect to the argument.
Note that this is still some way off a solution for $\phi$, which requires a
solution of the above equations for $\frac{\pd\phi}{\pd x},\ \frac{\pd\phi}{\pd
t}$ which may then in principle be integrated.
There is a nice class of explicit solutions;
\be
\phi=F(\alpha x+\beta t)+\gamma x +\delta t
\ee
where $F$ is arbitrary, and the constants $\alpha,\beta,\gamma,\delta$
satisfy the polynomial equation
\be
\lambda\alpha\beta+2\alpha\beta\gamma\delta-\beta^2\gamma^2-\alpha^2\delta^2=0.
\ee
The complete solution may be obtained in principle by the inversion of the
equations for $x,t$ in terms of $u_1,\ u_2$, and the subsequent integration of
the equations
\bea
\frac{\pd\phi}{\pd x}&=&\frac{1}{\sqrt{u_1}-\sqrt{u_2}} \nonumber \\
\frac{\pd\phi}{\pd t}&=&\frac{\sqrt{u_1u_2}}{\sqrt{u_1}-\sqrt{u_2}}
\label{ansa}
\eea
The consistency condition which guarantees integrability of these
equations is simply the Born Infeld equation itself. Thus in principle this
equation is as fully integrable as is the Bateman equation, even though
according to the analysis of \cite{fair}, it admits only a single Lagrangian,
namely $\sqrt{\lambda+4\frac{\pd\phi}{\pd x}\frac{\pd\phi}{\pd t}}$.

Here is an example of a solution generated by the results (\ref{hodo})
and (\ref{ansa}). A particularly convenient choice of $f$ and $g$ in
the system (\ref{hodo}) is
$f(u_1)=u_1^2$ and $g(u_2)=-u_2^2$. This gives explicit expressions
for $u_1$ and $u_2$ in terms of $x$ and $t$:
\bea
  u_1 & = & -\frac{(4x+t^2)}{4t} \nonumber \\
  u_2 & = & -\frac{(4x-t^2)}{4t}.
\eea
These can be substituted back into the formulae (\ref{ansa}) which can
then be integrated to give an explicit solution:
\be
\phi =-\frac{\sqrt{2}}{3}((\frac{4x-t^2}{-2t})^{\frac{3}{2}}+(\frac{4x
+t^2} {-2t})^{\frac{3}{2}}) + {\rm constant.}
\ee

\section{Legendre Transforms}
   The Legendre Transform, which was used in \cite{fab} to linearize the
Universal Field Equation may also be used to linearize (\ref{univ2}). This
transform, which is clearly involutive, has the flavour of a twistor transform.

The multivariable version of this transform runs as follows \cite{cour}.
Introduce a dual space with co-ordinates $\xi_i,\ i=1,\dots,d$ and a function
$w(\xi_i)$ defined by
\bea
\phi(x_1,x_2,\dots,x_d)+w(\xi_1,\xi_2,\dots,\xi_d)=
&x_1\xi_1+x_2\xi_2+\dots ,x_d\xi_d.\\
\xi_i={\pd\phi\over\pd x_i},\quad x_i={\pd w\over\pd \xi_i},\quad& \forall{i}.
\eea
To evaluate the second derivatives $\phi_{ij}$ in terms of derivatives of $w$
it is convenient to introduce two Hessian matrices;
$\Phi,\  W$ with matrix elements  $\phi_{ij}$ and $w_{\xi_i\xi_j}=w_{ij} $
respectively. Then assuming that $\Phi$ is invertible, $\Phi W=1\!\!1$
and
\be
 {\pd\sp2\phi\over\pd x_i\pd x_j}= ( W\sp{-1})_{ij},\quad
 {\pd\sp2w\over\pd \xi_i\pd \xi_j}= (\Phi\sp{-1})_{ij}.
\ee

 The effect of the Legendre transformation upon the equation (\ref{univ2}) is
immediate; in the new variables the equation becomes simply
\be
\sum_{i,j}(\xi_i\xi_j+f(\sum\xi_k^2)\eta_{ji}) \frac{\pd^2 w}{\pd \xi_i\pd
\xi_j}=0.
\label{univ3}
\ee
a linear second order equation for $w$. All sums are implicitly taken with the
Lorentzian metric. Introducing the variable
$\rho=\sqrt{\sum\xi_k^2}$ this equation takes the form

\be
[(\rho^2+f(\rho))\frac{\pd^2}{\pd\rho^2}+\frac{df(\rho)-1}{\rho}\frac{\pd}{\pd\rho}+
\frac{f(\rho)}{2\rho^2}\sum_{i,j}(\xi_i\frac{\pd}{\pd\xi_j}-\xi_j\frac{\pd}{\pd\xi_i})^2]w=0.
\label{dalembert}
\ee
Single valued solutions of this equation are eaily obtained when it is realised
that the eigenfunctions of the generalised total angular momentum  operator
$\sum_{i<j}(\xi_i\frac{\pd}{\pd\xi_j}-\xi_j\frac{\pd}{\pd\xi_i
})^2$ are just
harmonic functions on the $d-1$ sphere, with eigenvalues $-n(n+d-2)$, $n$
integral. Then the general solution can be found by the method of of separation
of variables as $w=\sum_nF_n(\rho)\times$(general harmonic of degree\ $n$),
where $F_n(\rho)$
is a solution to the ordinary differential equation
\be
(\rho^2+f(\rho))\frac{d^2F}{d\rho^2}+\frac{df(\rho)-1}{\rho}\frac{dF}{d\rho}-n(n+d-2)\frac{f(\rho)}{\rho^2}F=0.
\ee
Given such a solution, a parametric representation for $x_i$ in terms of
$\xi_j$
can be constructed from $x_i=\frac{\pd w}{\pd\xi_i}$, and these relations,
together with the definition of $w$ in terms of $\phi$ are sufficient to
eliminate the variables $\xi_j$ and solve for $\phi$. Of course this procedure
is only a solution in principle; in practice there will be comparatively few
solutions for $w$ for which an explicit solution can be obtained. It is clear
from  this construction that

\section{ Lagrangian Derivation}
In contrast to the Universal Field Equation  which possesses an infinite number
of inequivalent Lagrangian functions of which it is the resultant variation,
the modification (\ref{univ2}), admits only one (\cite{fair}). However this
Lagrangian retains one feature of the fully covariant situation; it may be
expressed in terms of an iteration of Euler variations.
Denote by ${\cal E}$ the Euler differential operator
\be
{\cal E}=-{\pd\over\pd\phi}
 +\pd_i {\pd\over\pd\phi_{x_i}}-\pd_i\pd_j{\pd\over\pd\phi_{{x_i},{x_j}}}\dots
\label{eul}
\ee
(In principle the expansion continues indefinitely  but it is sufficient here
 to terminate at the stage involving the  variational operator
$\frac{\pd}{\pd\phi_{{x_i}{x_j}}}$).
Now consider the Lagrangian density
\be
L_1= \sqrt{\sum_{i,j}\frac{\pd \phi}{\pd x_i}
         \frac{\pd \phi}{\pd x_j}\eta_{ij}+\lambda}
\ee
with equation of motion
\be
{\cal E}L_1=\frac{1}{L_1^{\frac{3}{2}}}\sum_{ij}
(L_1\eta_{ij}-\frac{\pd \phi}{\pd x_i }\frac{\pd \phi}{\pd x_j}) \frac{\pd^2
\phi}{\pd \xi_i\pd \xi_j}=0.
\label{n1}
\ee
where all the  contractions are taken using the Lorentz metric. In the case
where the dimension of space time is 2, (\ref{n1}) is simply the 2 dimensional
Born-Infeld equation (\ref{1}). Now relax the imposition of zero on the right
hand side of (\ref{n1}) and define a new Lagrangian density,
$ L_2=L_1{\cal E}L_1$, with equation of motion
\be
{\cal
E}L_2=\frac{1}{L_1^{\frac{3}{2}}}\sum_{ij}\sum_{kl}(L_1\eta_{ij}\eta_{kl}-\frac{\pd \phi}{\pd x_k }\frac{\pd \phi}{\pd x_j}\eta_{il}) (
\frac{\pd^2 \phi}{\pd \xi_i\pd \xi_j}\frac{\pd^2 \phi}{\pd \xi_k\pd
\xi_l}-\frac{\pd^2 \phi}{\pd \xi_i\pd \xi_k}\frac{\pd^2 \phi}{\pd \xi_j\pd
\xi_l})=0.
\ee
 In the case of a 3 dimensional space time this equation is just (\ref{univ2}).
The general pattern is clear; defining recursively the density $L_n$ by
$L_n=L_1{\cal E}L_{n-1}$, the equation in $d$ dimensions is simply given by
$L_{d-1}=0.$ This is in exact parallel with the iterative generation
of the Universal Field Equation, where the only difference is that instead of a
specific choice for $L_1$, any function of $\phi_{x_j}$  which is homogeneous
of weight one, with a vanishing Hessian $\det(\phi_{x_j,x_k})$ will do
\cite{fai}.

\section{Acknowledgement }

J.A. Mulvey acknowledges the support of the Department of Education
for Northern Ireland.

\end{document}